\documentstyle[12pt]{article}
\input psbox
\newcommand{\beq}{\begin{equation}}
\newcommand{\eeq}{\end{equation}}
\newcommand{\beqn}{\begin{eqnarray}}
\newcommand{\eeqn}{\end{eqnarray}}
\newcommand{\stackM}{\stackrel{\scriptstyle >}{{ }_{\sim}}}
\newcommand{\stackm}{\stackrel{\scriptstyle <}{{ }_{\sim}}} 

\newcommand{\ca}{c_\alpha}
\newcommand{\sa}{s_\alpha}
\newcommand{\sbe}{s_\beta}
\newcommand{\cb}{c_\beta}

\begin{document}

\thispagestyle{empty}

\begin{flushright}
{\parbox{3.5cm}{
UAB-FT-392

May, 1996

hep-ph/9605306
}}
\end{flushright}
\vspace{3cm}
\begin{center}
\begin{large}
\begin{bf}
MSSM AND Z WIDTH: IMPLICATIONS ON TOP QUARK AND HIGGS PHYSICS
\footnote{Invited talk presented at the XXXIst Rencontres de Moriond, 
{\it Electroweak Interactions and Unified Theories}, Les Arcs,
 March 1996. To appear in the Proceedings (Editions Fronti\`eres).}\\
\end{bf}
\end{large}
\vspace{1cm}
Joan SOL\`A\\

\vspace{0.25cm} 
Grup de F\'{\i}sica Te\`orica\\
and\\ 
Institut de F\'\i sica d'Altes Energies\\ 
\vspace{0.25cm} 
Universitat Aut\`onoma de Barcelona\\
08193 Bellaterra (Barcelona), Catalonia, Spain\\
\end{center}
\vspace{0.3cm}
\hyphenation{super-symme-tric}
\hyphenation{com-pe-ti-ti-ve}
\begin{center}
{\bf ABSTRACT}
\end{center}
\begin{quotation}
\noindent
We discuss the one-loop quantum corrections to the top quark
width and to the hadronic widths
of the Higgs bosons of the MSSM, and emphasize the results obtained
in particular regions of the MSSM parameter space which have been 
proposed to alleviate the anomalies observed in the
decay modes of the $Z$ into $b\bar{b}$ and  $c\bar{c}$. 
We find that the corrections can be large   
and should be visible through the measurement of the
top quark production cross-sections in future
experiments at the Tevatron and at the LHC.
\end{quotation}
 
\baselineskip=6.5mm  %(FOR PREPRINT)

\newpage

Lately we have been witnesses to a flood of experimental
information potentially challenging the predictions of the SM
to an unprecedented level. We are referring to the recent scrutinies of high
precision electroweak data
on $Z$-boson observables\,\cite{Hildreth}, where the ``anomalies'' 
in the ratios $R_b$ and $R_c$, far from disappearing, apparently have
consolidated their
status in the context of the SM.
On the face of it, one is tempted to  
believe that physics might be taking a definite trend beyond the SM.  
There is in the literature
a fairly big amount of early as well as of very recent SUSY work on
$R_b$\,\cite{Pokorski,GJS2}
showing that the discrepancies
can be significantly weakened.
 The fact that the anomalies in $R_b$ and $R_c$ --and 
perhaps also in $\alpha_s(M_Z^2)$\,\cite{Shifman}-- can be
simultaneously minimized\,\cite{GS12} 
after including supersymmetric quantum effects on the
$Z$-boson partial widths      
is a remarkable feature of the
Minimal Supersymmetric Standard Model (MSSM) which should not be understated. 

In view of the new wave of SUSY potentialities,
it is natural to study the possible
consequences that the particular subspace of MSSM parameters singled out
by the $Z$ observables may have in other areas of Particle Physics. As shown in
Refs.\cite{GJS2,GS12}, one privileged piece of parameter space, 
call it Region I, is characterized by a large value of $\tan\beta$
($\sim m_t/m_b$) in conjunction with 
a moderate value of both the CP-odd Higgs and the lightest chargino mass 
around $60\,GeV$.
In Refs.\,\cite{GJSH}-\cite{GJS}  
we studied the impact of Region I of the MSSM
parameter space on the top-quark width. The result for the standard decay
$t\rightarrow W^+\,b$ is that the overall (electroweak plus strong) SUSY 
corrections can be relatively large ($\stackM 10\%$) in some cases, but in
general the SUSY corrections to $t\rightarrow H^+\,b$ are larger and
in contradistiction to the former case remain
sizeable even for all sparticle masses well above $100\,GeV$. 
Another relevant region (Region II) is characterized 
by a large value of the CP-odd Higgs mass, $M_{A^0}$, namely of a few 
hundred $GeV$,
and by a relatively light chargino and stop of $60\,GeV$.
Here we encounter what we dubbed the ``tangential solution'' 
to the $R_b$ anomaly (see Fig.4 of Ref.\cite{GJS2}); in the light of the
present data on $R_b$\,\cite{Hildreth} it is no longer a
strict  ``solution'', though it
is certainly much better than the SM prediction, for it
only differs $-1.5\,\sigma$ from the central value of
$R_b^{\rm exp}=02211\pm 0.0016$, in comparison to the
more severe $-3.4\,\sigma$ discrepancy afflicting the SM prediction. 
In Region II the harmful (negative)
effects on $R_b$ from heavy higgses are restrained since $\tan\beta$ is 
assumed to lie in the intermediate interval $ 2\stackrel{\scriptstyle <}
{{ }_{\sim}} \tan\beta
\stackrel{\scriptstyle <}{{ }_{\sim}} 20$ (see Fig.3 of Ref.\cite{GJS2}),
so that the aforementioned ``tangential solution'' becomes optimized and
is entirely due to genuine (R-odd) supersymmetric particles. 
We remark that although Region I
is more efficient to mitigate the $R_b$ anomaly, its scope is limited to
$M_{A^0}<70\,GeV$ (equivalently,
$M_{H^+}\stackrel{\scriptstyle <}{{ }_{\sim}} 100\,GeV$).
In contrast, the ``tangential solution'' 
in Region II extends the range of  $M_{A^0}$ (and therefore that of $M_{H^+}$)
from about $100\,GeV$ up to about $1\,TeV$.
Finally, we may envision another situation (Region III) 
where the $R_b$ anomaly can also be alleviated   
by pure supersymmetric quantum effects, namely, it is
characterized by the same sparticle spectrum as in Region II together with very 
small values of $\tan\beta$ ($\stackrel{\scriptstyle <}{{ }_{\sim}}0.7$)
and very large (effectively decoupled) $M_{H^{\pm}}>1\,TeV$. 
This solution lies in the far left edge of
Fig.4 of Ref.\cite{GJS2} but it is not explicitly exhibited there. 
Since for our purposes we are not interested in a too heavy MSSM
Higgs spectrum, we shall not 
dwell on Region III any further. 

From the foregoing considerations, we see that the state of the art 
in $Z$-boson anomalies is such that
present-day quantum physics of $Z$-boson observables
still tolerates both the ``light'' ($M_{A^0}<m_t$) and heavy ($M_{A^0}>m_t$)
kinematical domains of MSSM Higgs boson masses. (Of course, the CP-even mass
remains always relatively light, $M_{h^0}\stackm 130\,GeV$.)
 Notwithstanding, these domains
are qualitatively very different. 
Whereas in Region I 
the charged Higgs decay of the
top quark, $t\rightarrow H^+\,b$, is the relevant process to deal
with\,\cite{GJS},
in Region II  $M_{H^+}$ can be sufficiently large to prompt 
the top-quark decay of a supersymmetric charged Higgs boson,
$H^+\rightarrow t\,\bar{b}$. In view of the possible existence of
Region II compatible with $Z$-boson data,
in the present note we shall be concerned with
the bearing of SUSY physics on that decay,
perhaps one of the most  important decay modes to search for at the 
LHC and also in the next generation of Tevatron experiments. 
The study of the quantum effects on $H^+\rightarrow t\,\bar{b}$ could be the
clue to unravel the potential supersymmetric nature of the charged Higgs.

Furthermore, should the physical domain of the MSSM
parameter space turn out to fall in Regions I or II, then we shall see that  
the hadronic widths
$\Gamma (A^0,h^0,H^0)\rightarrow q\bar{q}$ of the neutral Higgs bosons
of the MSSM
must, too, incorporate
important virtual SUSY signatures to look for
which can be extracted from measured quantities by subtracting  
the corresponding conventional QCD corrections\,\cite{QCD}.
In our analysis we will neglect those decays leading to
light $q\,\bar{q}$ final states since their branching ratios are very small.
Thus, for the lightest neutral Higgs, $h^0$, we will concentrate on just
the decay $h^0\rightarrow b\,\bar{b}$ (which can take place only
in Region I),
whereas for $A^0, H^0$ 
we shall consider the 
channels $A^0, H^0\rightarrow b\,\bar{b}$ and
$A^0, H^0\rightarrow t\,\bar{t}$ (involving Regions I/II and II, respectively)
with the understanding that the $t\,\bar{t}$ modes are dominant, if available 
(i.e. if $M_{A^0}>2\,m_t$).

The basic free parameters of our analysis concerning the electroweak sector
are contained in the
stop and sbottom mass matrices ($q=t,b$): 
\begin{equation}
{\cal M}_{\tilde{q}}^2 =\left(\begin{array}{cc}
M_{\tilde{q}_L}^2+m_q^2+\cos{2\beta}(T^3_q-
Q_q\,\sin^2\theta_W)\,M_Z^2 
 &  m_q\, M_{LR}^q\\
m_q\, M_{LR}^q &
M_{\tilde{q}_R}^2+m_q^2+Q_q\,\cos{2\beta}\,\sin^2\theta_W\,M_Z^2\,.  
\end{array} \right)\,,
\label{eq:stopmatrix}
\end{equation}
with 
\beq
M_{LR}^{\{t,b\}}=A_{\{t,b\}}-\mu\{\cot\beta,\tan\beta\}\,, \ \ \ \
\eeq
$\mu$ being the SUSY Higgs mixing parameter in the superpotential.
The $A_{t,b}$ are the trilinear soft SUSY-breaking parameters and the
$M_{{\tilde{q}}_{L,R}}$ are soft SUSY-breaking masses.
By $SU(2)_L$-gauge invariance we must have $M_{\tilde{t}_L}=M_{\tilde{b}_L}$,
whereas $M_{{\tilde{t}}_R}$, $M_{{\tilde{b}}_R}$ are in general independent
parameters. In the strong supersymmetric sector, the basic parameter is the
gluino mass, $m_{\tilde{g}}$.
For the sake of simplicity in the presentation, we shall treat the sbottom mass 
matrix by freezing the sbottom mixing parameter at $M_{LR}^b=0$ and assuming 
equal eigenvalues:
$m_{\tilde{b}_1}=m_{\tilde{b}_2}\equiv
m_{\tilde{b}}$. As far as 
the stop mass matrix is concerned it is generally non-diagonal
($M_{LR}^t\neq 0$) and the stop masses are different, with the
convention $m_{\tilde{t}_2}>m_{\tilde{t}_1}$.

For the full SUSY corrections to $t\rightarrow W^+\,b$ and 
$t\rightarrow H^+\,b$
we refer the reader to the detailed
Refs.\,\cite{GJSH,GJS,CGGJS}. Here we limit 
ourselves to report on the conventional QCD and SUSY-QCD contributions 
to the decays $\Gamma ( H^+\rightarrow t\,b)$ and
$\Gamma (A^0,h^0,H^0)\rightarrow q\bar{q}$. Indeed, the strong
corrections mediated by quarks, gluons and their SUSY partners
(squarks and gluinos, respectively) are expected to be the leading
corrections to these decays in the MSSM.
(The analysis of the larger and far more complex
body of SUSY-electroweak corrections to these modes, namely the
corrections mediated by squarks, sleptons, chargino-neutralinos and the 
Higgs bosons themselves, is given in Refs.\,\cite{CGGJS,Dabelstein}.)

To compute the one-loop SUSY-QCD corrections to
$\Gamma\equiv\Gamma (H^{+}\rightarrow t\, \bar{b})$ in the MSSM,
we shall adopt the on-shell renormalization scheme\,\cite{BSH} where the
fine structure constant,
$\alpha$, and the masses of the gauge bosons, fermions and scalars are
the renormalized parameters: $(\alpha, M_W, M_Z, M_H, m_f, M_{SUSY},...)$.
The interaction Lagrangian describing the $H^{\pm}\,t\,b$-vertex  
in the MSSM reads as follows: 
\beq
{\cal L}_{Htb}={g\,V_{tb}\over{2}M_W}\,H^+\,\bar{t}\,
[a_L^+(t)\,P_L + a_R^+(b)\,P_R]\,b+{\rm h.c.}\,,
\label{eq:LtbH}
\eeq 
where $P_{L,R}=1/2(1\mp\gamma_5)$ are the chiral projector operators,
$V_{tb}$ is the corresponding CKM matrix element--henceforth
we set $V_{tb}=1$-- and
\beq
a_L^+(t)=\sqrt{2}\,m_t\cot\beta\,,\ \ \ \
a_R^+(b)=\sqrt{2}\,m_b\tan\beta\,.
\label{eq:clR}
\eeq
Similarly, the interaction Lagrangian describing 
the various neutral Higgs decays $\Phi^i\rightarrow q\,\bar{q}$   
($\Phi^1\equiv A^0,\,\Phi^2\equiv h^0,\,\Phi^3\equiv H^0$) 
at tree-level in the MSSM
reads as follows: 
\beq
 {\cal L}_{\Phi q q}= \frac{g}{2 M_W}\Phi^i \bar{q} 
                       \left[ a_L^i(q) P_L +a_R^i(q) P_R\right] q\,.
\label{eq:LHqq}
\eeq 
We shall focus on top and bottom quarks ($q=t,b$). 
In a condensed and self-explaining notation we have defined
\beqn
a_R^1(t,b)&=&-a_L^1(t,b)= m_q (i\cot\beta,i\tan\beta)\,,\nonumber\\
a_R^2(t,b)&=&a_L^2(t,b)= m_q (-\ca/\sbe,\sa/\cb)\,,\nonumber\\
a_R^3(t,b)&=&a_L^3(t,b)= m_q (-\sa/\sbe,-\ca/\cb)\,,
\label{eq:aLaR}
\eeqn
with $\ca\equiv\cos\alpha, \sbe\equiv\sin\beta$ etc. (Angles $\alpha$ 
and $\beta$
are related in the usual manner as prescribed by the MSSM\,\cite{Hunter}).

The one-loop corrected amplitudes for all the decays above
have the generic form ($i=+,1,2,3$):
\beq
  iO^i = \frac{i g}{2 M_W}
        \left[ a_L^i(q)\left(1+O_L^i(q)\right) P_L +
          a_R^i(q')\left(1+O_R^i(q')\right) P_R\right]\,.
\eeq
The renormalized form factors read
\beqn
  O_L^i(q)&=&K_L^i(q)+\frac{\delta m_q}{m_q}+\frac{1}{2}\delta Z_L^q+
                                   \frac{1}{2}\delta Z_R^q\,,
  \nonumber\\
  O_R^i(q')&=&K_R^i(q')+\frac{\delta m_{q'}}{m_{q'}}
+\frac{1}{2}\delta Z_L^{q'}+ \frac{1}{2}\delta Z_R^{q'}\,,
\eeqn 
where the $K_{L,R}^i(q)$ stand for the $3$-point function contributions,
and the remaining terms include the mass and wave-function renormalization
in the on-shell scheme (see the similar analysis of Ref.\cite{GJS}). 

After explicit computation of the various loop diagrams, the results are 
conveniently casted in terms of the relative
correction with respect to the corresponding tree-level width, $\Gamma_0$:
\begin{equation}
\delta={\Gamma-\Gamma_0\over \Gamma_0}\,.
\label{eq:delta}
\end{equation}
A crucial parameter entering the loop contributions is $\tan\beta$.
In supersymmetric theories, like the MSSM, the spectrum of higgses and of
Yukawa couplings is richer than in the SM and, in such a framework, the
bottom-quark Yukawa coupling may counterbalance the smallesness of the bottom
mass at the expense of a large value of $\tan\beta$, the upshot being that
the top-quark and bottom-quark Yukawa couplings in the superpotential
read
\begin{equation}
h_t={g\,m_t\over \sqrt{2}\,M_W\,\sin{\beta}}\;\;\;\;\;,
\;\;\;\;\; h_b={g\,m_b\over \sqrt{2}\,M_W\,\cos{\beta}}\,.
\label{eq:Yukawas} 
\end{equation}
Thus, since the perturbative domain of these Yukawa couplings is
$0.5\stackm \tan\beta \stackm 70$, there is room enough
for both $h_t$ and
$h_b$ being larger than the gauge coupling, $g$, and even comparable
to one another (for $\tan\beta=m_t/m_b\simeq 35$).  This is so in Region I 
where $\tan\beta$ is very large, and is also partly the case in Region II
where $\tan\beta$ can be moderately high.

In Fig.1a we plot the width of $H^+\rightarrow t\,\bar{b}$
versus $\tan\beta$ after including 
SUSY-QCD effects and compare it with the corresponding tree-level
width, $\Gamma_0(H^+\rightarrow t\,\bar{b})$, as well as 
with the partial widths of two alternative (non-hadronic) modes.
We have fixed
$m_t=180\,GeV$, $M_{H^+}=250\,GeV=M_{LR}^t$,
the squark masses $m_{\tilde{t}_1}=60\,GeV$, $m_{\tilde{b}}=150\,GeV$, and
the gluino mass $m_{\tilde{g}}=200\,GeV$. It is patent that the corrections
can be very large in the high $\tan\beta$ regime. This can be further
appreciated in Fig.1b, where we plot both the SUSY-QCD ($\delta_{\tilde{g}}$)
and the conventional QCD ($\delta_g$) corrections (\ref{eq:delta}) against
$M_{H^+}$. It is seen that in the very relevant region
$M_{H^+}\leq 500\,GeV$ and
depending on the actual value 
of $\tan\beta$, the SUSY-QCD effects can be comparable or even be dominant over
the standard QCD corrections.
Since $\delta_{\tilde{g}}$ can have either sign (opposite to the sign of
$\mu$), the total QCD
correction in the MSSM,  $\delta_g+\delta_{\tilde{g}}$, could be much
smaller than what would be naively
expected within the context of conventional quantum chromodynamics. 
Alternatively, $\delta_g+\delta_{\tilde{g}}$ could be
extremely large near the $t\,\bar{b}$ threshold, suggesting that
in this case higher orders
should be taken into account. 
%\newpage
%%%%%%%%%%%%%%%%%%%%%%%%%%%%%%%%%%%%%%%%%%%%%%%%%%%%%%%%%%%%%%%%%%%%%%%
\hspace{-1cm}
\begin{figure}
%%%
%%%%%%  Figures Centrades Verticalment
%%%
\newcommand{\vcpsboxto}[3]{{$\vcenter{{\psboxto(#1;#2){#3}}}$}}
%%%
%       \vcpsboxto{x_dimension}
%                 {y_dimension}
%                 {name_of_encapsulated_postcript_file}
%%%

%\begin{tabular}{ll}
%\vspace{-4cm}
%\hspace{-1.8cm} \vcpsboxto{10cm}{0cm}{fig1.eps} &
%\hspace{-1.8cm} \vcpsboxto{10cm}{0cm}{fig2.eps} \\
%\end{tabular}

%\vspace{-1.8cm}
%\begin{center}
%\vcpsboxto{8cm}{0cm}{fig3.eps}
%\end{center}

\begin{center}
%{\bf Fig. 1}
\hspace*{-1cm}
\vcpsboxto{18cm}{0cm}{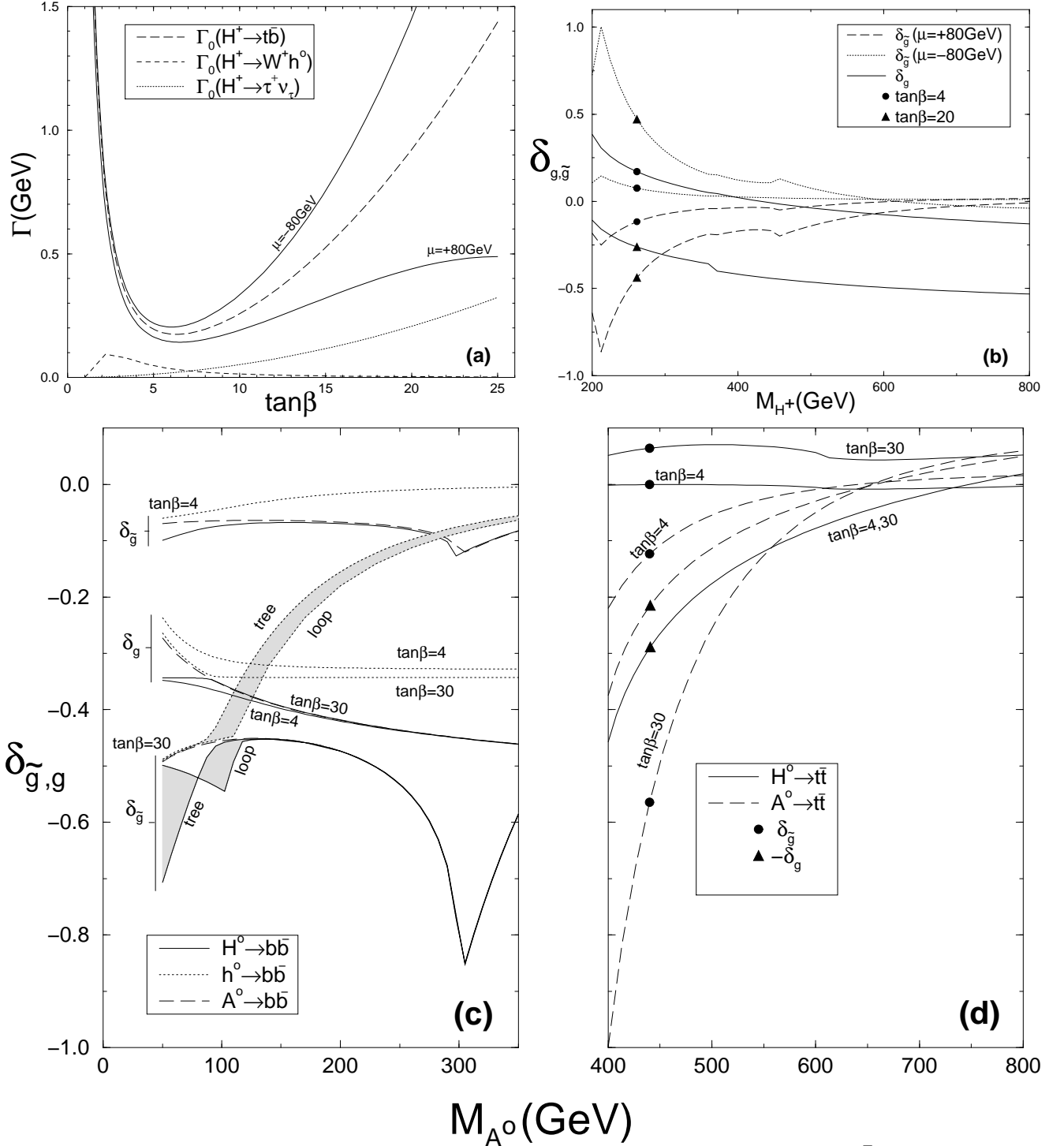}
\end{center}
%%\begin{center}
%\noindent

\vspace{-1.5cm}
\caption
{(a) The SUSY-QCD corrected and uncorrected width
$\Gamma (H^+\rightarrow t\,\bar{b})$ as
a function of $\tan\beta$ (remaining parameters given on the text)
as compared to two alternative (non-hadronic) decays; 
(b) The SUSY-QCD ($\delta_{\tilde{g}}$) and QCD ($\delta_g$) corrections
to $\Gamma (H^+\rightarrow t\,\bar{b})$ versus $M_{H^+}$;
(c)-(d) As in (b), but for the hadronic widths of the neutral MSSM higgses
as a function of $M_{A^0}$ .}
%%\end{center}
\end{figure}
%%%%%%%%%%%%%%%%%%%%%%%%%%%%%%%%%%%%%%%%%%%%%%%%%%%%%%%%%%%%%%%%%%%%%%%%%%%% 

The analysis of $\delta_{\tilde{g}}$ and $\delta_g$ (or $-\delta_g$) for
the neutral Higgs decay modes is exhibited in Figs.1c-1d. 
We have fixed the same values for the squark masses, gluino mass and
mixing parameters as in the previous figures.
The shaded areas in Fig.1c
signal the differences between the corrections
obtained using the tree-level and the one-loop 
Higgs mass relations.  Although $\delta_g$ for neutral
higgses is strictly independent
of $\tan\beta$, for fixed Higgs masses, the indirect dependence is due to the 
evolution of the MSSM Higgs masses 
$M_{h^0}, M_{H^0}$ as a function of $M_{A^0}$ since this evolution does 
depend on $\tan\beta$.
In Regions I and II we see from Fig.1c that
the decays $h^0\rightarrow b\,\bar{b}$ and
$A^0\rightarrow b\,\bar{b}$ receive negative standard QCD corrections of about
$-35\%$.
 For $M_{A^0}\stackrel{\scriptstyle >}{{ }_{\sim}} 100\,GeV$,
the standard QCD correction to $h^0\rightarrow b\,\bar{b}$
remains saturated at that value since 
the mass $M_{h^0}$ also saturates at its maximum value, whereas the modes 
$A^0, H^0\rightarrow b\,\bar{b}$ obtain increasing negative corrections.  
In contrast, $A^0\rightarrow t\,\bar{t}$ and $H^0\rightarrow  t\,\bar{t}$,
receive positive standard QCD corrections
(in this case we have plotted $-\delta_{\tilde{g}}$ in Fig.1d), which are 
of order $40-50\%$ near the threshold.
It is clear from Figs. 1c-1d that in many cases the SUSY 
effects are important since
they can be of the same order as the standard QCD corrections. As a matter
of fact, there are situations where $|\delta_{\tilde{g}}|>|\delta_g|$, with
the possibility of the SUSY effects either reinforcing
the standard QCD corrections or cancelling them out, perhaps to the extend
of reversing their sign!. Obviously, this feature could be used to 
differentiate SUSY and non-SUSY neutral 
higgses produced in the colliders.

In summary, we have
found that $t\rightarrow H^+\,b$ or $ H^+\rightarrow t\,\bar{b}$ and/or
$A^0, H^0\rightarrow b\,\bar{b}, t\,\bar{t}$
could be the ideal experimental
environment where to study the nature of the spontaneously symmetry
breaking mechanism. It could even 
be the right place where to target our long and unsuccessful 
search for large, {\it and} slowly decoupling,
quantum supersymmetric effects. In this respect it should be
stressed the fact that the typical size of our corrections is maintained 
even for all sparticle masses well above the LEP $200$ discovery range. 
Fortunately,
the next generation of experiments at the Tevatron and the future high precision
experiments at the LHC may well acquire the ability to test the kind of effects
considered here\,\cite{Willenbrock}. 
In fact,  since the leading Higgs production vertices are the same as the
hadronic Higgs decay vertices that we have been dealing with
in the present work, we expect that $t$ and $t\bar{t}$
can be copiously produced in association with MSSM Higgs particles
in Regions I and II (where the SUSY Yukawa
couplings $h_t$ and $h_b$ are enhanced).
Moreover, in these regions the large SUSY corrections reported here
automatically 
translate into sizeable effects on the cross-sections,
a fact which may constitute a distinctive imprint of SUSY.

As a final remark, we emphasize that the vigorous quantum
effects ($\sim 50\%$) potentially underlying the dynamics of the MSSM 
higgses are in stark contrast to the milder SUSY 
radiative effects (few per mil!) expected in the physics of
gauge bosons\,\cite{GJS2}. Needless to say, the excellence of the latters
over the formers lies in
the high precision techniques achieved by the $Z$ experiments
at LEP, which might enable us to resolve
the small SUSY corrections and perhaps confirm that they are responsible 
for the anomalies observed in the $Z$ width.
Be as it may, our analysis suggests that
Higgs physics at the colliders might well take its turn in the near future
and  eventually  
provide the most straightforward handle to ``virtual'' Supersymmetry.

%%%%%%%%%%%%%%%%%%%%%%%%%%%%%%%%%%%%%%%%%%%%%%%%%%%%%%%%%%%%%%%%%%%%%%%%%%%%%%
%\vspace{0.6cm}
{\bf Acknowledgements}:

\noindent
The author is thankful to R.A. Jim\'enez and J.A. Coarasa for their
collaboration
in the preparation of this contribution.
This work has been partially supported by CICYT 
under project No. AEN93-0474. 

\baselineskip=4.8mm
%\newpage


\begin{thebibliography}{9999}
\bibitem{Hildreth}
M. Hildreth, these proceedings.
\bibitem{Pokorski}
S. Pokorski, these proceedings;
M. Boulware, D. Finnell,\, {\it Phys. Rev.} {\bf D 44} (1991) 2054; 
J.D. Wells, C. Kolda, G.L. Kane,\, {\it Phys. Lett.} {\bf B 338} (1994) 219.
\bibitem{GJS2}
D. Garcia, R.A. Jim\'enez, J. Sol\`a,\, {\it Phys. Lett.} 
{\bf B 347} (1995) 321 [E: {\bf B 351} (1995) 602].
\bibitem{Shifman}
M. Shifman, these proceedings.
\bibitem{GS12}
D. Garcia, J. Sol\`a,\, {\it Phys. Lett.}\, {\bf B 354} (1995) 335 and
{\bf B 357} (1995) 349.
\bibitem{GJSH}
D. Garcia, W. Hollik, R.A. Jim\'enez and J. Sol\`a, {\it Nucl. Phys.}
{\bf B 427} (1994) 53;
A. Dabelstein, W. Hollik, R.A. Jim\'enez, C. J\"unger,
J. Sol\`a,\, {\it Nucl. Phys.}\, {\bf B 456} (1995) 75.
\bibitem{GJS}
J. Guasch, R.A. Jim\'enez, J. Sol\`a,\,{\it Phys. Lett.}\, {\bf B 360} (1995) 47.
\bibitem{QCD}
C.S. Li, R. Oakes, {\it Phys. Rev.}\,{\bf D 43} (1991) 855;
E. Braaten, J.P. Leveille, {\it Phys. Rev.}\,{\bf D 22} (1980) 715. 
\bibitem{CGGJS}
J. A. Coarasa, D. Garcia, J. Guasch, R.A. Jim\'enez, 
J. Sol\`a,\, preprint UAB-FT (in preparation).
\bibitem{Dabelstein}
A. Dabelstein, W. Hollik, preprint MPI-Ph.93-86;
A. Dabelstein,\,preprint KA-THEP-4-1995. 
\bibitem{BSH}
M. B\"ohm, H. Spiesberger, W. Hollik,\, {\it Fortschr. Phys.}
{\bf 34} (1986) 687;
W. Hollik,\,  {\it Fortschr. Phys.} {\bf 38} (1990) 165.  
\bibitem{Hunter}
 J.F. Gunion, H.E. Haber, G.L. Kane, S. Dawson,\, {\it The Higgs Hunters's
Guide} (Addison-Wesley, Menlo-Park, 1990).
\bibitem{Willenbrock}
S. Willenbrock, these proceedings; C.-P. Yuan, preprint hep-ph/9604434.



\end{thebibliography}
\end{document}